\documentclass[12pt]{article}
\usepackage{amssymb,amsmath,epsfig}

\begin{document}

\title{\bf Dark Energy Models and Laws of Thermodynamics in Bianchi I Model}
\author{M. Sharif \thanks {msharif.math@pu.edu.pk} and Rabia Saleem
\thanks{rabiasaleem1988@yahoo.com}\\
Department of Mathematics, University of the Punjab,\\
Quaid-e-Azam Campus, Lahore-54590, Pakistan.}

\date{}
\maketitle

\begin{abstract}
This paper is devoted to check validity of the laws of
thermodynamics for LRS Bianchi type I universe model which is filled
with combination of dark matter and dark energy. We take two types
of dark energy models, i.e., generalized holographic dark energy and
generalized Ricci dark energy. It is proved that the first and
generalized second law of thermodynamics are valid on the apparent
horizon for both the models. Further, we take fixed radius $L$ of
the apparent horizon with original holographic or Ricci dark energy.
We conclude that the first and generalized second laws of
thermodynamics do not hold on the horizon of fixed radius $L$ for
both the models.
\end{abstract}

\section{Introduction}

Our universe has a phase transition from decelerating to
accelerating. Type Ia supernova \cite{1}-\cite{5} indicates that the
universe has accelerated expansion, i.e., the universe is expanding
with accelerating velocity. The main reason of this expansion is
said to be a mysterious energy with large negative pressure known as
dark energy (DE). Recent data \cite{6}-\cite{8} shows that DE
occupies $76\%$ of the universe and $24\%$ is occupied by some other
matter. The cosmological constant is the best identification of this
mysterious energy, but it raises some other issues like fine-tuning
and cosmic-coincidence puzzle. The equation of state parameter
(EoS), $\omega=-1$ is the most acceptable candidate for DE. Some
dynamical models also help to understand its nature like
quintessence \cite{9,10}, K-essence \cite{11}, tachyon field
\cite{12,13}, Chaplygin gas \cite{14,15}, phantom model
\cite{16}-\cite{18} etc.

Recent literature indicate keen interest to check the validity of
first and second law of thermodynamics. In different scenarios, Wang
et al. \cite{19} investigated that these laws are valid on the
apparent horizon, when different from the event horizon, but do not
hold for event horizon. Mazumder and Chakraborty \cite{20,21}
explored some conditions for the validity of generalized second law
of thermodynamics (GSLT) by using the first law of thermodynamics.
Debnath \cite{22} investigated validity of GSLT by using holographic
DE (HDE) interacting with two fluids for FRW model. Mubasher et al.
\cite{23} proved that GSLT holds for all time and does not depend
upon geometry and EoS parameter. The validity of GSLT has also been
checked on Kaluza-Klein cosmology with modified HDE (MHDE)
\cite{24}. In a recent paper \cite{25}, the validity of GSLT is
investigated when anisotropic DE is interacting with anisotropic
radiations and DM in BI universe model.

Holographic DE principle is also used to study the nature of DE.
This principle states that in a bounded system the number of degrees
of freedom should not be infinite and system is scaled by its
boundary area but not by its volume \cite{26}. Cohen and his
collaborators \cite{27} explained the relationship between short
distance cutoff $\Lambda$ and long distance cutoff $L$ by
considering the quantum field theory. They proposed a limiting
energy bound, a system with size $L$ cannot form black hole if the
vacuum energy of the system exceeds than its mass of the same size
$L$. This can be written as $L^3\rho_{\Lambda}\leq LM_{p}^{2}$,
where $\rho_\Lambda$ is the quantum zero point energy density, $L$
is infrared cutoff and $M_p$ is the reduced Planck mass expressed as
$M_P=(8\pi G)^{-\frac{1}{2}}$. This inequality is possible only for
large $L$, so the HDE density can be expressed as
$\rho_\Lambda=3c^2M_p^{2}L^-2$, where $3c^2$ is a dimensionless
constant.

Ricci DE (RDE) \cite{28} is a type of DE obtained by taking square
root of the inverse Ricci scalar as its infrared cutoff. Gao et al.
\cite{29} explored that the DE is proportional to the Ricci scalar.
Some recent work \cite{30}-\cite{33} shows that the RDE model fits
well with observational data. Xu et al. \cite{34} defined two types
of DE models, i.e., generalized HDE (GHDE) and generalized RDE
(GRDE) models

In this paper, we use LRS BI universe model composed of DM and DE
with GHDE and GRDE models. The paper is organized as follows: In
section \textbf{2}, the density and pressure for GHDE and GRDE
models are found. Section \textbf{3} is devoted to check the
validity of the first and GSLT on the apparent horizon and also by
taking GHDE or GRDE as the original HDE or RDE. In the last section,
we conclude the results.

\section{Density and Pressure for GHDE and GRDE models}

In this section, we formulate the field equations for LRS BI
universe model. We then evaluate density and pressure for GHDE as
well as GRDE models. The line element of LRS BI model is given as
follows
\begin{equation}\label{1}
ds^2=-dt^2+A^2(t)dx^2+B^2(t)(dy^2+dz^2),
\end{equation}
where $A(t)$ and $B(t)$ are scale factors. We use the well-known
condition $A=B^m$ \cite{35}-\cite{38}, where $m$ is a positive
constant. Consequently, the above metric reduces to
\begin{equation}\label{2}
ds^2=-dt^2+B^{2m}(t)dx^2+B^2(t)(dy^2+dz^2).
\end{equation}
The first field equation corresponding to this metric gives us
\begin{eqnarray}\label{3}
(2m+1)\frac{\dot{B}^2}{B^2}=8\pi \rho,\\\label{4}
H_2^{2}=\frac{8\pi}{1+2m}\rho,
\end{eqnarray}
where $H_2=\frac{\dot{B}}{B}$ is the directional Hubble parameter.
The conservation equation yields
\begin{equation}
\dot{\rho}+(m+2)H_2(\rho+P)=0.\label{5}
\end{equation}
where $\rho$ and $p$ are the energy density and pressure of the
fluid, respectively. Taking derivative of Eq.(\ref{4}) and using
Eq.(\ref{5}), it follows that
\begin{equation}\label{6}
\dot{H}_2=-\frac{4\pi(m+2)}{1+2m}(\rho+P).
\end{equation}

We consider that the fluid is a combination of DM and DE, i.e.,
$\rho=\rho_m+\rho_{DE}$ and $P=P_m+P_{DE}$ with $P_m=0$. Assuming
that there is no interaction between DM and DE so that these are
separately conserved. Thus the conservation equation (\ref{6}) gives
\begin{eqnarray}\label{7}
\dot{\rho}_m+2(m+2)\rho_mH_2=0,\\\label{8}
\dot{\rho}_{DE}+(m+2)(\rho_{DE}+P_{DE})H_2=0.
\end{eqnarray}
Equation (\ref{7}) gives the energy density of matter as
\begin{equation}\label{9}
\rho_m=\rho_{m_0}(1+z)^3,
\end{equation}
where $\rho_{m_0}$ is an integration constant which gives the
present value of DE density and $z$ is the red shift given as
\begin{equation*}
z=\frac{1}{B^{\frac{m+2}{3}}}-1.
\end{equation*}
In the following, we evaluate energy density and pressure for GHDE
and GRDE models.

\subsection{Generalized Holographic Dark Energy Model}

The energy density of GHDE model is given by \cite{33}
\begin{equation}\label{10}
\rho_h=\rho_{DE}=\frac{3c^2}{8\pi}H^2f(\frac{R}{H^2}),
\end{equation}
where $H=\frac{1}{3}(m+2)\frac{\dot{B}}{B}$, $c (\neq0)$ is an
arbitrary constant and $f(x)>0$ such that $f(x)=\alpha
x+(1-\alpha),~0\leq\alpha\leq1$. The Ricci scalar is given by
\begin{equation}\label{11}
R=-2\left[(m^2+2m+3)H_{2}^{2}+(m+2)\dot{H}_2\right].
\end{equation}
When $\alpha=0$, the energy density of the original HDE is
recovered, while for $\alpha=1$, we obtain energy density of the
original RDE. Using Eq.(\ref{11}) in (\ref{10}), it follows that
\begin{equation}\label{12}
\rho_h=\frac{c^2}{24\pi}\left[-18(m+2)\alpha\dot{H}_2+[(m+2)^2-(19m^2+40m+58)\alpha]
H_{2}^{2}\right].
\end{equation}
Inserting Eqs.(\ref{9}) and (\ref{12}) in (\ref{4}), we obtain first
order linear differential equation whose solution is
\begin{equation}\label{13}
H_{2}^{2}=\frac{2}{24\pi\rho_{m_0}}\gamma(1+z)^3+H_{0}^{2}(1+z)^{\beta},
\end{equation}
where
\begin{eqnarray}\nonumber
\beta&=&\frac{1+2m}{2(m+2)^2\alpha
c^2}\left[1-\frac{c^2}{3(1+2m)}[(m+2)^2-(19m^2+40m+58)\alpha]\right],\\\nonumber
\gamma&=&3(1+2m)-[(m+2)^2-(m^2-32m-14)\alpha]c^2,
\end{eqnarray}
$H_0$ is an integration constant. Differentiating Eq.(\ref{13}) with
respect to $t$, we get
\begin{equation}\label{14}
\dot{H}_2=-\frac{12\pi\rho_{m_0}(m+2)}{\gamma}(1+z)^3-H_{0}^{2}\frac{\beta}{6(m+2)}
(1+z)^{\beta}.
\end{equation}
Substituting $H_{2}^2$ and $\dot{H}_2$ in Eqs.(\ref{11}), (\ref{12})
and (\ref{8}), we obtain the Ricci scalar, density $\rho_h$ and
pressure $P_h$, respectively
\begin{eqnarray}\label{15}
&&R=\frac{-24\pi(m^2+2)\rho_{m_0}}{\gamma}(1+z)^3
-\frac{H_{0}^{2}}{18\alpha c^2}\nonumber\\
&&\times\left[[(m+2)^2+(17m^2+32m+50)\alpha]c^2-3(1+2m)\right]
(1+z)^{\beta},\\\nonumber
&&\rho_h=\frac{[(m+2)^2-(10m^2+4m+22)\alpha]\rho_{m_0}c^2}{\gamma}(1+z)^3\\\label{16}
&&+\frac{H_{0}^{2}(1+z)^\beta}{48\pi}
\left[3(1+2m)+[(m+2)^2-(19m^2+40m+58)\alpha]c^2\right].\\\nonumber
&&P_h=\frac{H_{0}^{2}\gamma(1+z)^\beta}{864\pi(m+2)^2\alpha c^2}
\left[3(1+2m)-[(m+2)^2-(m^2-32m-14)\alpha]c^2\right].\\\label{17}
\end{eqnarray}
Equation (\ref{16}) and (\ref{17}) represent the density and
pressure in the form of red shift $z$.

\subsection{Generalized Ricci Dark Energy Model}

The energy density of GRDE model is given by \cite{33}
\begin{equation}\label{19}
\rho_r=\frac{3c^2}{8\pi}R
g(\frac{\frac{1}{9}(m+2)^2H_{2}^{2}}{R}),
\end{equation}
where $g(y)=\beta y+(1-\beta)>0,~0\leq\beta\leq1$. For $\beta=0$, we
recover energy density of the original RDE whereas $\beta=1$ leads
to energy density of the original HDE. Comparing Eqs.(\ref{11}) and
(\ref{19}), we see that the GRDE reduces to the GHDE and vice versa
for $\beta=1-\alpha$. If we replace $\alpha$ by $(1-\beta)$ in
Eqs.(\ref{13})-(\ref{17}), we obtain similar solutions for GRDE
model. This implies that these equations are also solutions of the
GRDE model with $\alpha=1-\beta$.

\section{First and Generalized Second Law of Thermodynamics}

First we discuss validity of the first and GSLT in $BI$ universe
bounded by apparent horizon. For this purpose, we use entropy given
by Gibb's law \cite{39,40}
\begin{equation}\label{20}
T_AdS_I=p dV+d(E_A),
\end{equation}
where $S_I,~V,~p,~E_A$ and $T_A$ are internal entropy, volume,
pressure, internal energy within the apparent horizon and
temperature of the apparent horizon, respectively. The internal
energy is $E_A=\rho V$ and  $V=B^{m+2}$ while the radius of the
apparent horizon in the case of flat geometry coincides with Hubble
horizon is given by
\begin{equation}\label{21}
R_A=L=\frac{1}{H}=\frac{3}{(m+2)H_2}.
\end{equation}
We assumed that the system is in equilibrium so that the fluid and
the horizon has the same temperature. The temperature and entropy of
the apparent horizon are defined as \cite{41}
\begin{eqnarray}\label{22}
T_A=\frac{1}{2\pi L},\quad S_A=S_h=\kappa \pi L^2.
\end{eqnarray}
The entropy of the horizon is $S_h=\frac{\kappa A^*}{4},~\kappa$ can
be taken $1$ in energy units, $A^*=4\pi L^2$ is the area of black
hole.

Now we check validity of the first law of thermodynamics on the
apparent horizon given by
\begin{equation}\label{24}
-dE_A=T_AdS_A.
\end{equation}
The energy crossing on the apparent horizon for BI universe model
can be found as follows \cite{42}
\begin{eqnarray}\label{23}
-dE_A&=&\frac{4\pi}{3}(m+2)L^3H_2T_{\mu
\nu}K^{\mu}K^{\nu}dt=\frac{8\pi}{3}(m+2)L^3H_2\rho dt
\nonumber\\&=&-\frac{(1+2m)}{3}H_2\dot{H}_2L^3dt.
\end{eqnarray}
Inserting the value of $L$ from Eq.(\ref{21}) in this equation, we
obtain
\begin{equation}\label{25}
-dE_A=-\frac{9(1+2m)}{(m+2)^3}
\left(\frac{\dot{H}_2}{H_{2}^{2}}\right)dt,
\end{equation}
Also, we have
\begin{equation}\label{26}
T_AdS_A=\dot{L}dt=
-\frac{3}{(m+2)}\left(\frac{\dot{H}_2}{H_{2}^{2}}\right)dt.
\end{equation}
These two equations lead to
\begin{equation}\label{27}
-dE_A=kT_AdS_A,
\end{equation}
where $k=\frac{3(1+2m)}{(m+2)^2}$. This shows that the first law of
thermodynamics always holds on the apparent horizon for all kinds of
energies as it is independent of DE.

For the validity of GSLT on the apparent horizon, we evaluate the
rate of change of internal entropy from Eq.(\ref{20}) as follows
\begin{equation}\label{27}
\dot{S}_I=\frac{(\rho+P)\dot{V}+V\dot{\rho}}{T_A}.
\end{equation}
Substituting the values of $\dot{V},~T_A$ and $\dot{\rho}$, it
follows that $\dot{S_I}=0$. According to SLT, entropy of the
thermodynamical system can never be decreased. This is generalized
in such a way that the derivative of any entropy is always
increasing, i.e., $\dot{S}_I+\dot{S}_A\geq 0$. Thus we have
\begin{eqnarray}\label{28}
\dot{S}_I+\dot{S}_A=
-\frac{18\pi}{(m+2)^2}\left(\frac{\dot{H}_2}{H_{2}^{2}}\right)dt
\geq 0.
\end{eqnarray}
This shows that the GSLT is always satisfied on the apparent
horizon. Notice that these laws always hold independent of choice of
DE on the apparent horizon.

Now we take GHDE or GRDE model as an original density of HDE and RDE
to check validity of the first and GSLT on the horizon of radius
$L$. The original HDE or RDE density is given by
\begin{equation}\label{29}
\rho_\Lambda=\frac{3c^2}{8\pi}L^{-2}.
\end{equation}
Comparing this value with the energy density of HDE (\ref{11}), it
follows that
\begin{equation}\label{30}
L^2=\frac{1}{\alpha R+\frac{1}{9}(1-\alpha)(m+2)^2H_{2}^{2}}.
\end{equation}
Substituting the values of $R$ and $H_2^2$ from Eqs.(\ref{12}) and
(\ref{14}), we can write this in the form of red shift
\begin{equation}\label{31}
L^2=\frac{18c^2\gamma}{48\pi\rho_{m_0}c^2\delta(1+z)^3+H_{0}^{2}\gamma\mu(1+z)^\beta},
\end{equation}
where
\begin{eqnarray*}
\delta&=&[(m+2)^2-2(5m^2+2m+11)\alpha],\\
\mu&=&[(m+2)^2-(19m^2+40m+58)\alpha]c^2+3(1+2m).
\end{eqnarray*}

Here the temperature and entropy on the horizon are similar to
Eq.(\ref{22}), i.e., $T_L=\frac{1}{2\pi L},~S_L=\pi L^2$. The amount
of energy crossing on the horizon is also similar to Eq.(\ref{23})
with the difference that $dE_L$ is written instead of $dE_A$. We can
write
\begin{equation}\label{34}
T_L dS_L=\dot{L}dt.
\end{equation}
For the first law, we must have $-dE_L=T_LdS_L$. Equations
(\ref{23}) and (\ref{34}) imply that
\begin{equation}\label{35}
-dE_L=T_LdS_L-\left[\frac{(1+2m)}{3}H_2\dot{H}_2L^3+\dot{L}\right]dt.
\end{equation}
Since the second term on right hand side is time dependent, so
during evolution of the universe, it can never be zero, hence
\begin{equation}\nonumber
-dE_L \neq T_L dS_L.
\end{equation}
This indicates that the first law of thermodynamics does not hold on
the horizon of radius $L$ in Einstein's gravity.

For the validity of GSLT on the horizon of radius $L$, the rate of
change of total entropy is
\begin{equation}\label{37}
\dot{S}_I+\dot{S}_L=2\pi L\dot{L}.
\end{equation}
\begin{figure}
\epsfig{file=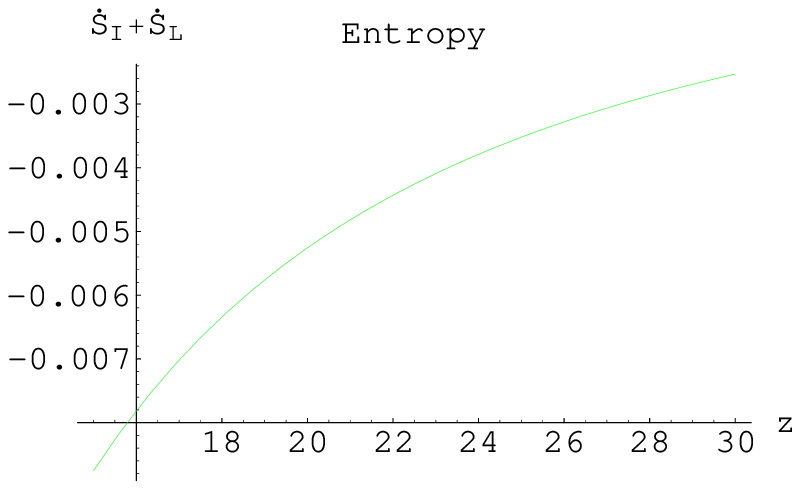, width=0.45\linewidth}\epsfig{file=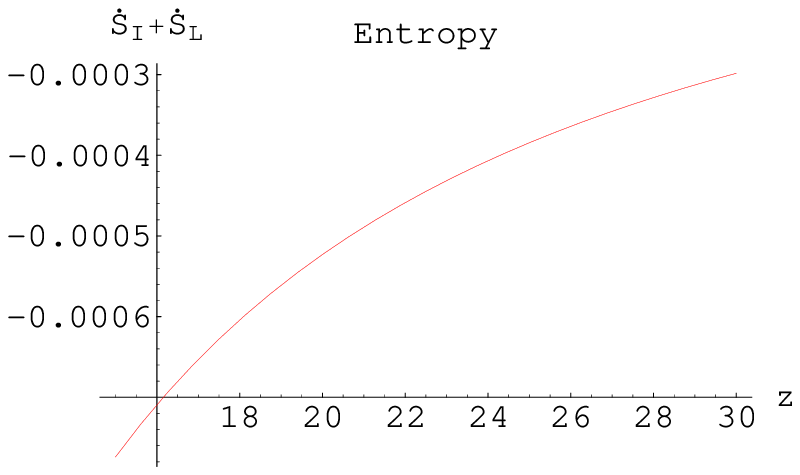,
width=0.45\linewidth} \caption{The graph shows the variation of
$(\dot{S}_I+\dot{S}_L)$ against red shift $z$ for
$c=0.5,~\rho_{m_0}=1,~\alpha=0.7,~\beta=0.7,~H_0=70$. The green
colour represents GHDE and red represents GRDE.}
\end{figure}
According to the GSLT, the total entropy of the thermodynamical
system always increases, i.e., $L\dot{L}\geq 0$ indicating its
dependence only on $L$ in the DE model. In GHDE model, the rate of
change of total entropy on the horizon is
\begin{eqnarray}\label{38}
\dot{S}_I+\dot{S}_L=-\frac{(m+2)\pi(1+z)}{3}\frac{dL^2}{dz}
\left[\frac{24\pi\rho_{m_0}}{\gamma}(1+z)^3+H_{0}^{2}(1+z)^{\beta}\right]^\frac{1}{2},
\end{eqnarray}
where $L^2$ is given in Eq.(\ref{31}). This is very complicated
expression in $z$ which does not provide any indication whether it
increases or decreases. To get insight, we draw the total entropy
$(\dot{S}_I+\dot{S}_L)$ versus red shift $z$ as shown in Figure
\textbf{1}. The graph indicates that $(\dot{S}_I+\dot{S}_L)<0$ and
hence the GSLT does not hold on the horizon of radius $L$. The green
and red lines represent the GHDE and GRDE models respectively.
Consequently, the GSLT does not hold for both kinds of the energy
models.

\section{Concluding Remarks}

In this paper, we have considered LRS BI universe model by assuming
that our universe is filled with DM and DE. We have taken two types
of DE models, i.e., GHDE and GRDE models. Notice that the GRDE model
can be converted to GHDE model if we replace $\beta$ by $1-\alpha$.
Also, the original density of HDE and RDE models is obtained for
$\alpha=0,~\beta=1$ and $\alpha=1,~\beta=0$ respectively. The
density and pressure for GHDE and GRDE models in terms of red shift
$z$ are evaluated.

The main purpose of this paper is to investigate validity of the
first and GSLT on the apparent horizon in this scenario. It turns
out that these laws are independent of the choice of DE models and
hence hold for any kind of DE models on the apparent horizon.
Further, we have considered the GHDE and GRDE as the original HDE
and RDE and found $L$ to check validity of these laws on this
horizon of radius $L$. It is found that the first and GSLT do not
hold on this horizon for both DE models. We would like to mention
here that in a recent work \cite{33}, similar type of investigation
has been done in FRW universe model. We have extended this work to
LRS BI universe model with same scenario. Here we can check validity
of the laws only on apparent horizon due to the flat geometry rather
than particle and event horizons as in the case of FRW universe.

\end{document}